\documentclass[%
 reprint,
nofootinbib,
 amsmath,amssymb,
 aps,
]{revtex4-2}

\usepackage{graphicx}
\usepackage{dcolumn}
\usepackage{bm}
\usepackage{siunitx}
\usepackage{multirow}
\usepackage{array,booktabs}

\begin{document}

\preprint{APS/123-QED}

\title{Machine Learning Force Field\\for Thermal Oxidation of Silicon}

\author{Lukas Cvitkovich}%
 \email{cvitkovich@iue.tuwien.ac.at}
\author{Franz Fehringer}%
\affiliation{%
 Institute for Microelectronics, Technische Universit\"at Wien, Austria
}%

\author{Christoph Wilhelmer}%
 \affiliation{%
 Institute for Microelectronics, Technische Universit\"at Wien, Austria
}%

\author{Diego Milardovich}%
 \affiliation{%
 Institute for Microelectronics, Technische Universit\"at Wien, Austria
}%

\author{Dominic Waldh\"or}%
\affiliation{%
 Institute for Microelectronics, Technische Universit\"at Wien, Austria
}%

\author{Tibor Grasser}%
\affiliation{%
 Institute for Microelectronics, Technische Universit\"at Wien, Austria
}%

\date{\today}

\begin{abstract}
Looking back at seven decades of highly extensive application in the semiconductor industry, silicon and its native oxide SiO$_2$ are still at the heart of several technological developments.
Recently, the fabrication of ultra-thin oxide layers has become essential for keeping up with trends in down-scaling of nanoelectronic devices and for the realization of novel device technologies.
With this comes a need for better understanding of the atomic configuration at the Si/SiO$_2$ interface.
Classical force fields offer flexible application and relatively low computational costs, however, suffer from limited accuracy.
Ab-initio methods give much better results but are extremely costly.
Machine learning force fields (MLFF) offer the possibility to combine the benefits of both worlds.
We train a MLFF for the simulation of the dry thermal oxidation process of a Si substrate. The training data is generated by density functional theory calculations. The obtained structures are in line with ab-initio simulations as well as with experimental observations. Compared to a classical force field, the most recent reactive force field (reaxFF), the resulting configurations are vastly improved. Our potential is publicly available in an open-access repository.


\end{abstract}

\maketitle


\section{\label{sec:intro}Introduction
}

Silicon has played a major role in semiconducting device technology for more than half a century and continues to find a broad range of novel applications spanning from single-electron devices~\cite{Guo1997} to spintronics~\cite{Jansen2012, Zutic_spintronics_review_2004} and semiconductor spin qubits~\cite{Zwanenburg2013}.

One of the most important reasons for the extensive use of Si is that its native oxide SiO$_2$ allows the production of semiconductor/insulator interfaces of exceptional quality~\cite{Razeghi2009}. 
Highly optimized devices like MOSFETs benefit from low defect densities at the interface and convenient growth of the oxide directly onto a Si substrate by thermal oxidation~\cite{Pantelides2006}.Although pure SiO$_2$ is being gradually substituted as a gate dielectric by other materials possessing significantly higher dielectric constants~\cite{Waldrop2016, 2D2020}, commonly termed as high-$k$ dielectrics, the inclusion of an ultra-thin SiO$_2$ passivation layer on the Si substrate prior to the application of the high-$k$ film remains crucial. In this regard, the SiO$_2$ passivation layer greatly enhances device performance, making it an essential component also in modern devices~\cite{HighK, HighK2, Nitride}.
Contemporary trends in fabrication and down-scaling of device dimensions have redirected research focus towards chemical-based bottom-up fabrication methods~\cite{NUR2020, HfO2onSi/Ge, lowTOx2007}. Among these methods, the creation of ultra-thin SiO$_2$ layers holds paramount significance.
Furthermore, as a playground for innovative device technologies, the material system Si/SiO$_2$ offers an appropriate environment for long-lived spins that can be controlled coherently~\cite{Veldhorst2014, cvitkovich_hf_2024}.

Ultra-thin layers of SiO$_2$ (on the order of a few nm) are typically fabricated through thermal oxidation of silicon. The underlying mechanisms of this process have been examined extensively over decades, through both experimental and theoretical means~\cite{DealGrove, Bongiorno2005, Bongiorno2004, Bongiorno2004_2, Pasquarello1998, interface1988, AKIYAMA2005, Gusev1995, Rosencher1979}. Earlier modeling approaches, like the seminal Deal-Grove model~\cite{DealGrove}, yield good results in a progressed stage of oxidation ($>15$\,nm oxide thickness), however, fail to describe the initial oxidation regime~\cite{Hopper1975, AFM2009} which is important for state-of-the-art technologies that require layered materials with thicknesses on the order of a few nm. Consequently, the intricacies of the initial oxidation phase requires more sophisticated models. In a previous study, based on the results of dynamic ab-initio molecular dynamics (AIMD) calculations, we proposed a multi-stage oxidation scheme that combined all previous experimental and theoretical insights into a comprehensive model~\cite{Cvitko2023}. 

\begin{figure*}[tp]
\centering
\includegraphics[width=\linewidth]{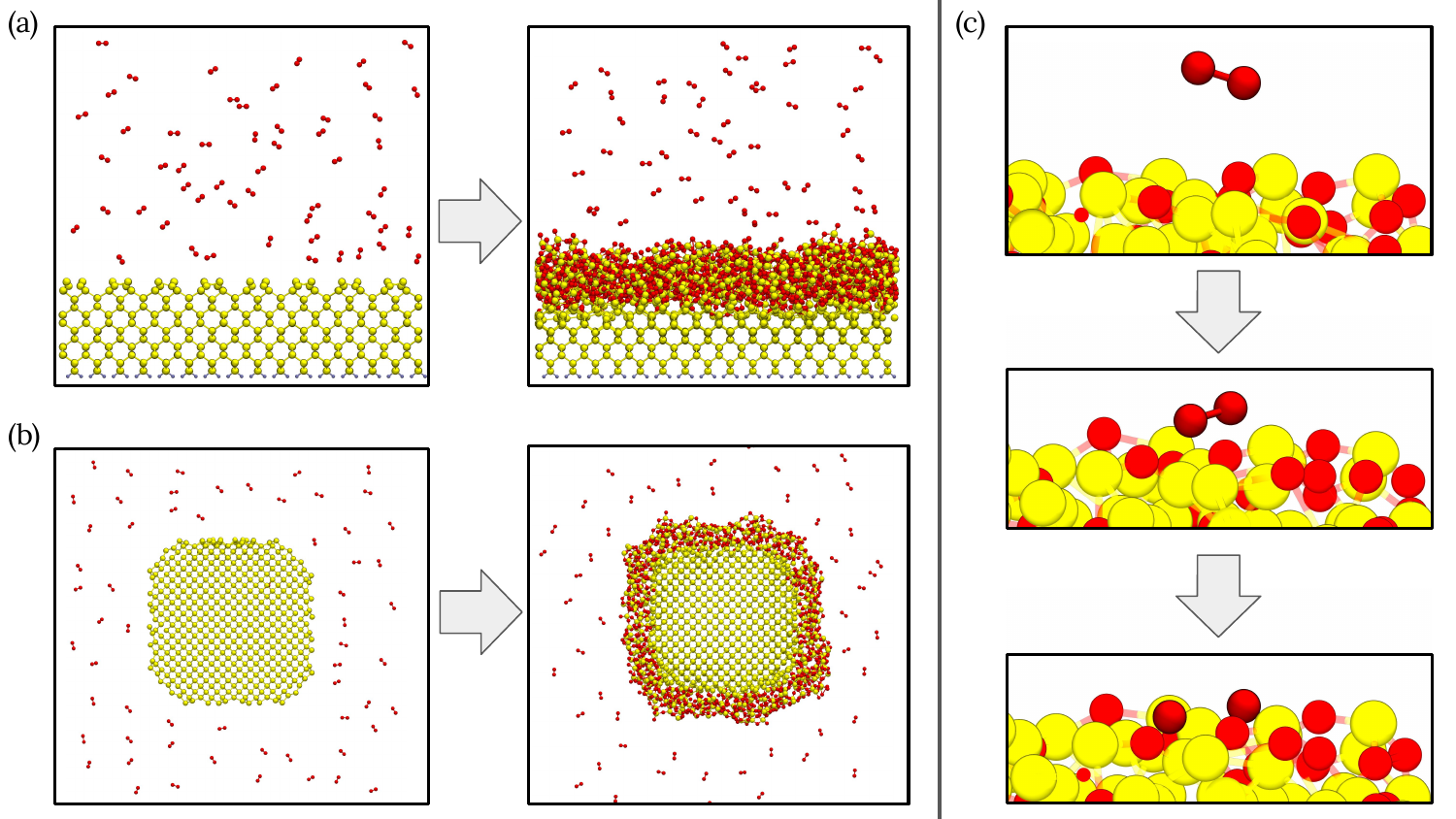}
\caption{Simulation procedure. \textbf{(a)} Reconfigurated Si(100) surfaces or \textbf{(b)} Si nanowires are exposed to gaseous oxygen within molecular dynamics calculations. \textbf{(c)} The O$_2$ molecules spontaneously react with the surface and dissociate to form an oxide coating around the surface. After every 10.000 steps ($\Delta t = 1$\,fs), the structure is optimized to avoid heating (each dissociation event releases energy), and O$_2$ molecules in the gas phase are refilled to maintain the pressure according to an ideal gas law.}
\label{fig:ML} 
\end{figure*}

In the present work, we extend our first-principles based modeling approach by a machine learning force field (MLFF). From the methodological perspective, the process of thermal oxidation offers an interesting use case for machine learned interatomic potentials, as our MLFF is universally suitable for gaseous oxygen, crystalline Si, and amorphous SiO$_2$. Machine learning techniques enable overcoming the strong limitations on cell sizes and simulation times, the typical drawbacks of ab-initio calculations, while keeping the accuracy of the results practically unchanged~\cite{Unke2021}. 
The ability to enlarge the system size and for simulating on larger time scales
allows for the investigation of the growth kinetics of the oxidation process and for the generation of even more realistic models of Si/SiO$_2$ interfaces, including long-range disorder and interface roughness. 
Furthermore, we extend our investigations from flat Si surfaces to more complex surface structures such as cylindrical Si nanowires. As shown in Fig.~\ref{fig:ML}, our ML approach allows modeling of the thermal oxidation process within dynamic simulations starting from entirely oxygen-free silicon surface structures. Within the MD simulation, these structures are exposed to an O$_2$ gas which reacts with the surface and forms a coating layer of amorphous SiO$_2$.

\section{Methodology}
Training of an MLFF typically requires the combined use of descriptors~\cite{SOAP}, machine learning algorithms~\cite{Bartok2010}, and high-accuracy training data. Methods and computational techniques employed in this work are described in the following.

~\subsection{Machine Learning Force Field}
Our MLFF is implemented within the Gaussian approximation potential (GAP) method~\cite{Bartok2010}. 
Similar to other ML models employed in the context of interatomic potentials, GAP completely neglects the electronic structure of a given system and assumes that the potential energy can be determined solely from the atomic configuration.
For this, GAP represents the total potential energy as the sum of local energy contributions from each atom. These local contributions are obtained by comparing a given local atomic environment to the local atomic environments of the atomic configurations available in the training dataset. The training dataset contains a number of atomic systems along with the corresponding energies that are taken from any other computational method that can be used to calculate the potential energy of an atomic system. GAP learns to estimate the energy of local atomic environments in the training process based on this data. Analogous to this procedure, GAP can also learn and estimate forces which further reduces the amount of necessary structures in the training dataset.

The usage of unbiased metrics is essential for the training data of a ML model. For this purpose, so-called descriptors are used to generate abstract representations of the input structures. A descriptor maps the atomic structure to a mathematical object (typically a vector) and this description is invariant to rotation, translation, or the permutation of identical atoms. This approach allows to reduce the size of the training dataset drastically by only providing the essential information to the ML model.
Within the scope of this work, we employed the frequently used smooth overlap of atomic positions (SOAP) descriptor~\cite{SOAP} in conjunction with two-body and three-body descriptors~\cite{two_three_body}. User-defined parameters are given in App.~\ref{app:descriptor}.

Our trained MLFF will be made available in an open-access repository, together with the training dataset~\cite{Download}.

~\subsection{DFT training data}

Our MLFF is trained entirely on data from more than 900 density functional theory (DFT) calculations. Underlying structures include single atoms, dimers, bulk structures of Si and SiO$_2$, as well as oxidized Si surfaces and nanowires with various oxygen coverage. 
Data for the initial training set is obtained by the stepwise oxidation process as presented in~\cite{Cvitko2023}.
In this approach, the starting points are oxygen-free Si surface structures which become gradually oxidized within AIMD calculations. Oxygen is provided by placing O$_2$ molecules in the vicinity of the Si surface. 
A comprehensive overview of the structures contained in the training dataset together with a detailed description of their generation can be found in App.~\ref{app:trainingdata}.

All density functional theory calculations are carried out using the CP2K package~\cite{Quickstep}, a code that uses the mixed Gaussian and plane waves approach (GPW). We use double-$\zeta$ Gaussian basis set for all atom types and the well-established Goedecker-Teter-Hutter (GTH) pseudopotentials to represent closed-shell electrons~\cite{BasisSet, Goedecker1996}. 
The electron density is expanded using a plane-wave basis with a cutoff of \SI{650}{Ry}.
The exchange-correlation energy is obtained by means of the semilocal generalized gradient approximation (GGA) functional PBE. 
Atomic relaxations were carried out with a force convergence criterion of 
$\SI[per-mode=repeated-symbol]{15}{\milli\electronvolt\per\angstrom}$.
Within the AIMD simulations, the total energy was conserved (microcanonical or NVE ensemble) and the total spin was restricted to $0$.

\section{Results}

\subsection{Comparison to DFT}
\begin{figure}[t]
\centering
\includegraphics[width=\linewidth]{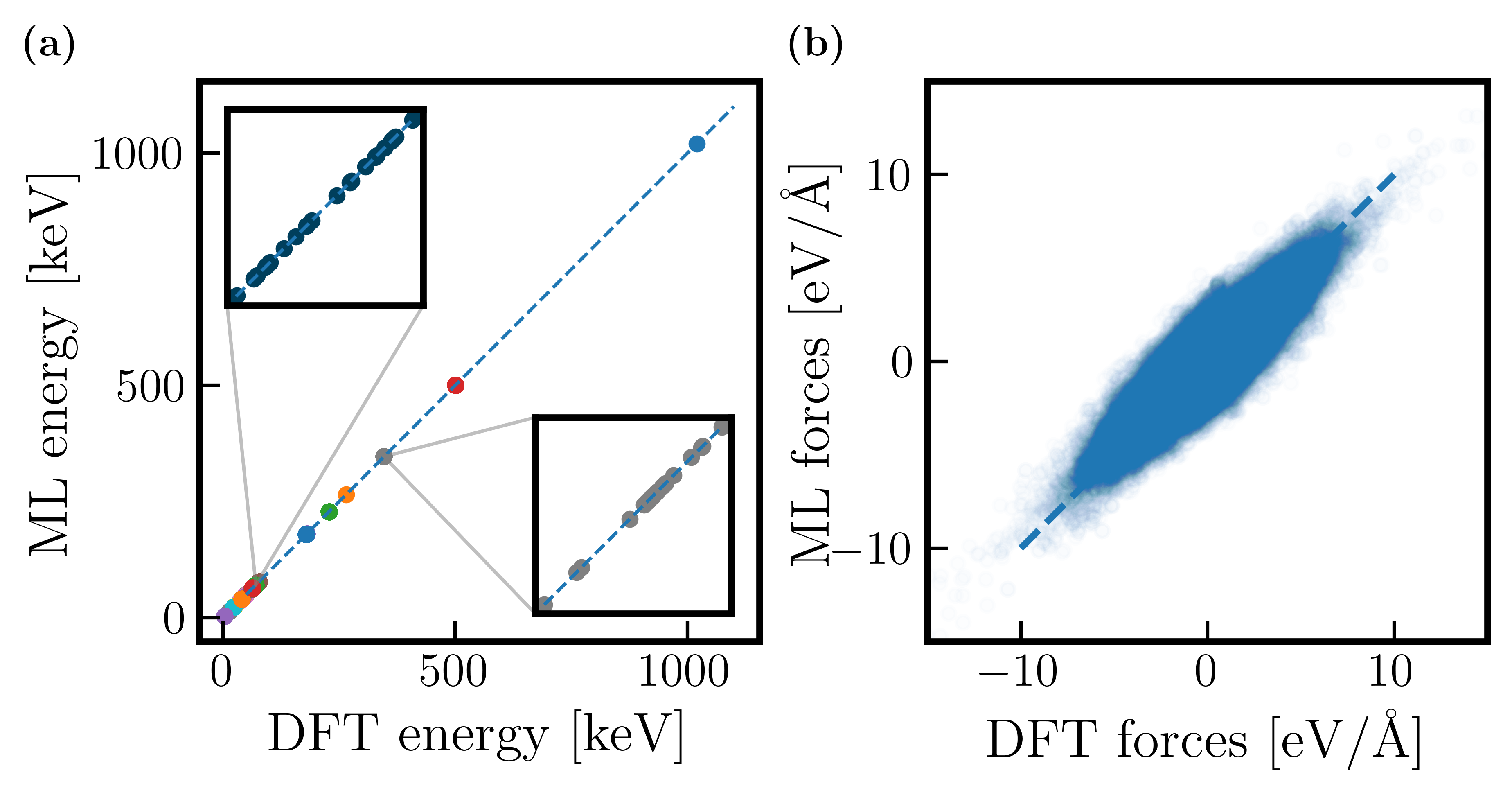}
\caption{Comparison between DFT and the ML model for more than 400 Si/oxide structures.  \textbf{(a)} The values of the energies show excellent correlation over the full range of structures as indicated by dashed blue lines with slope 1. Blue and grey points shown in the insets correspond to oxidized Si(100) surfaces and oxidized nanowires, respectively. The mean absolut errors (MAE) is below 10\,meV/atom for all tested structures. \textbf{(b)} Good agreement between the two methods is also obtained for the inter-atomic forces with an MAE below 0.16\,eV/\AA\,.
}
\label{fig:compare_energies}
\end{figure}

Validation of the ML model is done on a set of structures which are similar to the training dataset, i.e. structures of oxidized surfaces and nanowires, however, the structures are generated within MD runs with the ML model itself and subsequently recalculated by DFT. We compare the energies and forces predicted by DFT with the values from the ML model, see Fig.~\ref{fig:compare_energies}. The deviations are estimated by the mean absolute error (MAE) between both methods. The errors in energies are below 10\,meV/atom for all test structures, and the forces are predicted with an accuracy of 0.16\,eV/\AA. The clear linear correlation for systems with 200 to 5000 atoms indicates a very good agreement between DFT and the MLFF and allows to rule out systematic errors between the two.

\subsection{Structural properties}
\begin{figure}[t]
\centering
\includegraphics[width=\linewidth]{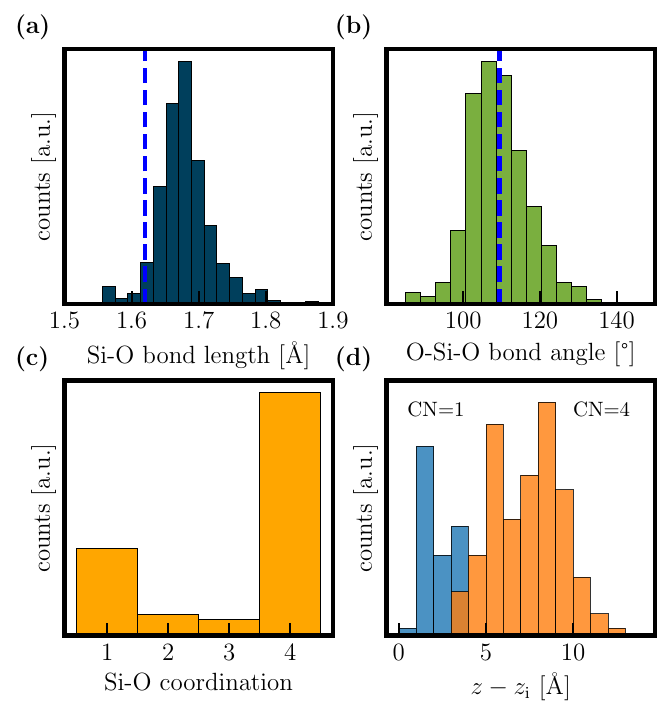}%
\caption{Geometric measures indicating the quality of the Si/SiO$_2$ structures. The results are shown for a representative interface structure with an effective oxide thickness of 1\,nm. \textbf{(a)} The Si-O bond lengths agree well with experimental values (dashed line) of 1.62\,$\mathrm{\AA}$ of bulk SiO$_2$~\cite{Mozzi1969, Diebold1999}. \textbf{(b)} Similar agreement is found for the O--Si--O bond angles which match the optimal tetrahedral bond angle of 109.47° (dashed line). \textbf{(c)} The O-coordination for Si atoms ranges from 1 to 4, as expected for interfacial structures. \textbf{(d)} Position of Si atoms with one O neighbor (CN=1) and four O neighbors (CN=4). Fourfold O-coordinated Si are found in SiO$_4$ tetrahedrons in the oxide while lower coordinations correspond to Si atoms close to the interface.
}
\label{fig:geometry}
\end{figure}

The ML model can be further validated by the structural qualities of the resulting structures, as shown in Fig.~\ref{fig:geometry}. The results refer to an oxidized Si surface which measures 6$\times $6\,nm$^2$ in plane and exhibits an oxide thickness of around 1\,nm.
While the mean Si--O bond length of 1.63\,\AA\, is in line with experimental values of bulk SiO$_2$~\cite{Diebold1999, Mozzi1969}, there are a number of strained bonds with a length of more than 1.8\,\AA. These bonds are exclusively found at the interface indicating a considerable strain in this region. 

An important result is the formation of SiO$_4$ tetrahedrons, the building blocks of SiO$_2$, indicating that even ultra-thin oxide layers already exhibit the structural properties of bulk SiO$_2$~\cite{Cvitko2023}. The mean O--Si--O bond angle (Fig.~\ref{fig:geometry}b) matches the ideal tetrahedral bond angle of \SI{109.47}{\degree} which means that the tetrahedrons are rigid and form already in an early stage of oxidation. The tendency to find enlarged O--Si--O bond angles (the angles between two tetrahedrons) at the interface (up to 140°) agrees with the previous observations of such interface structures~\cite{Hirose1999, Cvitko2023}. 
Further evidence for the formation of SiO$_2$ is provided by a coordination number analysis, see Fig.~\ref{fig:geometry}c and Fig.~\ref{fig:geometry}d. Most of the Si atoms in the oxide are fourfold coordinated by oxygen. Lower O-coordination can only be found at Si atoms close to the Si/oxide interface.
The interface is not sharp, but represented by a transition region of 0.5\,nm thickness. Within this region, the amount of oxygen increases steadily such that the O-coordination of the Si atoms increases from 0 to 4 in growth direction. Above the transition region, all Si atoms are four-fold O coordinated and integrated in a SiO$_4$ tetrahedron.
These results are not only in line with the AIMD simulations from~\cite{Cvitko2023}, they also agree with transmission electron microscope (TEM) images~\cite{Diebold1999, TEM2}, electron-energy-loss spectroscopy (EELS)~\cite{Muller1999, Muller2001}, and photoemission studies~\cite{Diebold1999, Photo2}.

\subsection{Growth kinetics}
As found experimentally~\cite{Hopper1975} and confirmed theoretically by means of AIMD calculations~\cite{Cvitko2023}, the oxidation rate decreases strongly as soon as an oxide layer has formed on the initially clean Si surface. As long as the Si surface is only sparsely oxidized, that is, the surface still shows unoxidized Si dimers, O$_2$ molecules can spontaneously adsorb and dissociate at the surface. During this phase, the oxidation rate is limited only by the amount of oxygen interacting with the surface.

\begin{figure}[t]
\centering
\includegraphics[width=\linewidth]{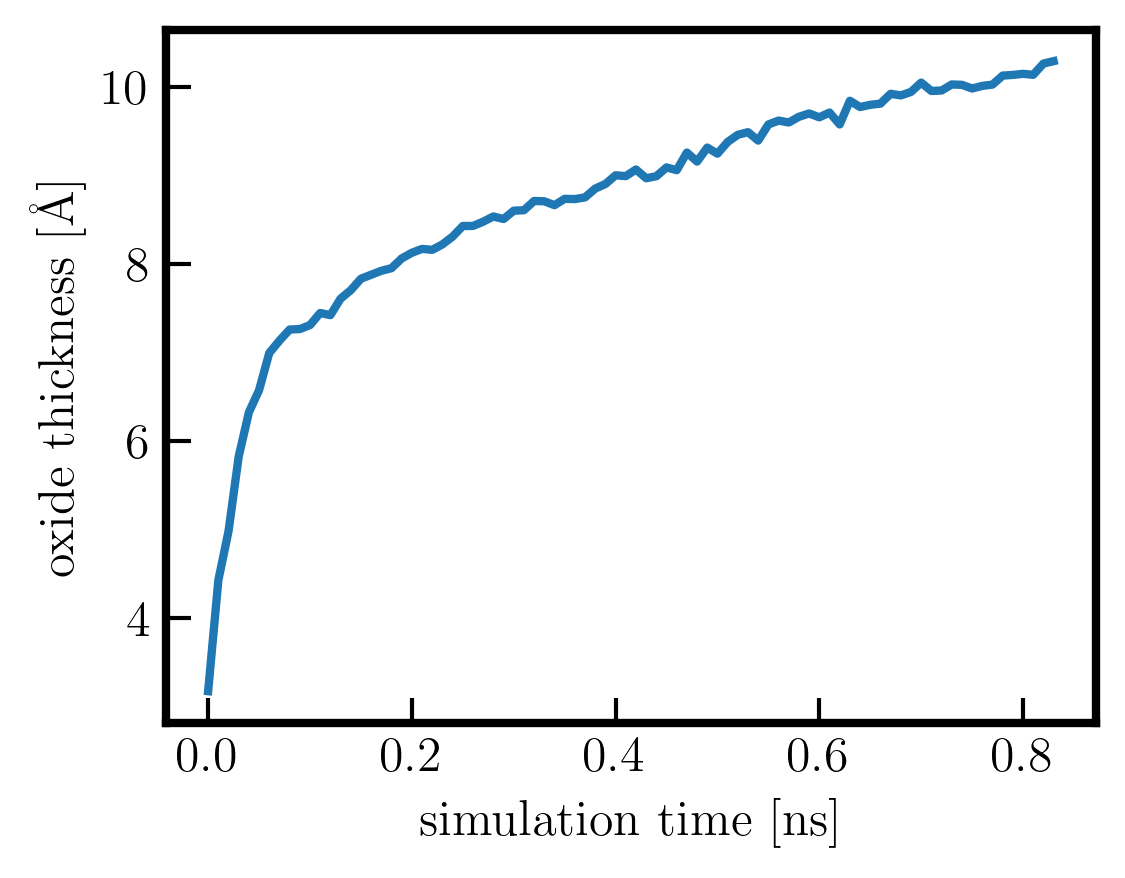}
\caption{Oxide thickness during dynamic oxidation of Si as a function of simulation time. Initially, fast oxidation is enabled by spontaneous surface reactions. After the surface is saturated with O, the dominant reaction mechanism changes to molecular precursor mediated dissociation, a process associated with slower oxidation rates~\cite{Liao2006}.}
\label{fig:ox_rate}
\end{figure}

In a later stage, in which the surface is fully covered by an oxide film, the oxidation rate decreases as the limiting factor is now the diffusion of O into deeper layers of Si. The diffusion is necessary in order to make room for further dissociative surface reactions. This behavior is also captured by the MLFF. We evaluate the position of the Si/SiO$_2$ interface $z_\mathrm{i}$ and oxide surface $z_\mathrm{s}$ by averaging the $z$-position of the five lowest and the five highest oxygen atoms, respectively. The oxide thickness $t$ is then the difference between surface and interface $t=z_\mathrm{i}-z_\mathrm{s}$. Evaluating the thickness of the oxide layer by this procedure in an MD run that simulates the thermal oxidation starting from a clean Si surface allows to estimate $t$ as a function of time as shown in Fig.~\ref{fig:ox_rate}.\footnote{We artificially enhance the growth rate by exposing the Si surface to a large number of O$_2$ molecules, corresponding to a pressure in the O$_2$ gas of $p=50$\,bar. This is necessary as the actual oxidation time (in the range of seconds) is still well outside the scope of feasible calculations even when using the MLFF.}
The oxidation rate has a maximum at the beginning and decreases significantly as soon as the surface is saturated with oxygen. At this point, O$_2$ molecules can not dissociate spontaneously anymore but adsorb onto the surface where they eventually dissociate after a few ps. This behavior explains the experimentally observed decrease in oxidation rate~\cite{Hopper1975} and is in line with previous observations from~\cite{Cvitko2023, Liao2006, O2_chemisorption_2023, Schubert_1993}.

\subsection{Interface quality}
Numerous experiments have shown that the growth of SiO$_2$ on a Si substrate results in a significant interface and surface roughness~\cite{Carim1987, Ohsawa2009}. In the initial oxidation regime, the roughness increases with the oxide thickness but saturates after the oxide layer exceeds 10\,nm. 
At this point, the oxidation rate becomes constant and the process is governed by O$_2$ diffusion~\cite{Ohsawa2009}, as assumed within the Deal-Grove model~\cite{DealGrove}, instead of O$_2$ surface reactions which enable a faster oxidation in the early oxidation stages~\cite{Cvitko2023}.

\begin{figure}[t]
\centering
\includegraphics[width=\linewidth]{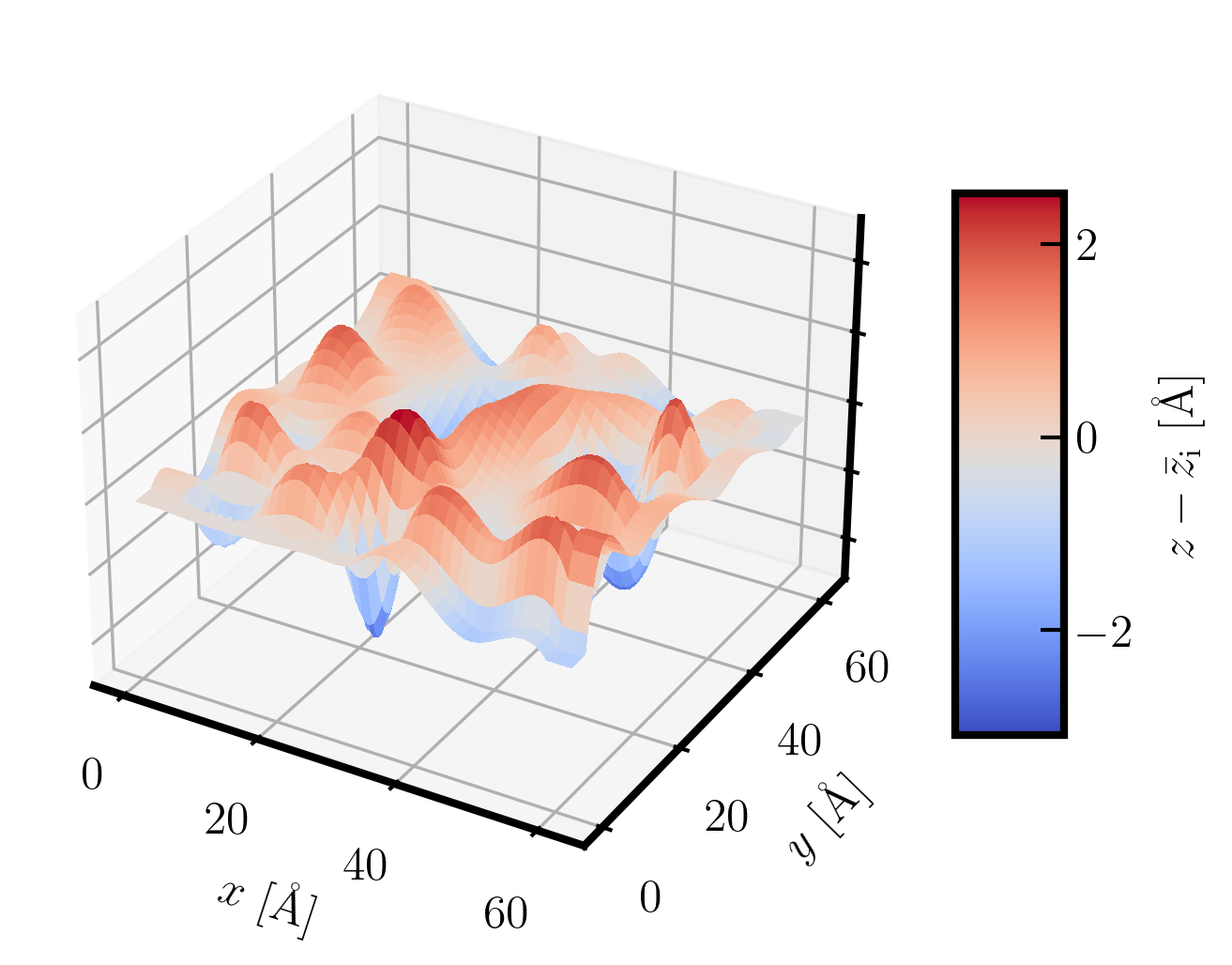}
\caption{Interface roughness as resulting due to the dynamic oxidation process governed by random adsorption trajectories. The RMS roughness of the interface is found to be $R_q=0.79\mathrm{\AA}$.}
\label{fig:roughness}
\end{figure}

In order to investigate the interface roughness, we oxidize a 6$\times6$\,nm$^2$ Si surface by means of the MLFF and depict the roughness of one representative Si/SiO$_2$ interface resolved in the in-plane directions in Fig.~\ref{fig:roughness}. For this analysis, we take the z-position of the lowest oxygen atom in each lateral 2D bin and connect their coordinates. The interface deviates from the average interface position $\bar{z}_\mathrm{i}$ by up to 2.5\,\AA{} with an average deviation of $R_a$=0.57\,\AA. Typically, the interface roughness is characterized by the root mean square deviations for which we find $R_q=0.79\,\mathrm{\AA}$ in reasonable agreement with the measured values reported in~\cite{Ohsawa2009}.

Furthermore, we find a mass density of around 2.5 g/cm$^3$ in the oxide layer which slightly overestimates experimental values of ultra-thin a-SiO$_2$ reported in the range of 2.24 to 2.36 g/cm$^{-3}$~\cite{Diebold1999}.
On the other hand, the mass density in the interface region complies with the experimental values from~\cite{Diebold1999}. 

\subsection{Comparison to classical force field}
In order to further validate our force-field, we compare it to one of the most commonly used classical force fields, namely the reactive force field (reaxFF) from~\cite{Nayir2019}.
A comparative data set is generated by dynamically oxidizing a $3\times3$\,nm$^2$ Si surface with the same initial parameters of position and velocity from identical starting configurations. The simulation time is set to 1.4\,ns integrated over 1.4 million time steps. As before, after every 10k steps, new oxygen molecules are added to the vacuum above the surface, to keep the pressure at 50\,bar.

Analyzing a number of geometric properties gives results summarized in Tab.~\ref{comparison}. In terms of two- and three-body geometric measures, our MLFF performs slightly better than the reaxFF potential. While the mean O--Si--O bond angles are close to the ideal tetrahedral angle, the reactive force field gives mean Si--O bond lengths of 1.55\,\AA\,(compared to the MLFF value of 1.68\,\AA\, and the experimental value of 1.62\,\AA~\cite{Diebold1999, Mozzi1969}). 
Furthermore, we compare the volumetric mass density in the interface region $\rho_\mathrm{IF}$ and in the oxide layer $\rho_\mathrm{OX}$. 
As mentioned before, our MLFF slightly overestimates the density in the oxide layer but reproduces the density in the interfacial transition region. ReaxFF on the other hand, gives densities about 10\% larger than experimental values~\cite{Diebold1999}, in line with implications from shortened bond lengths, in both the interface and the oxide regions of the interface structure. 

\begin{figure}[t]
\centering
\includegraphics[width=\linewidth]{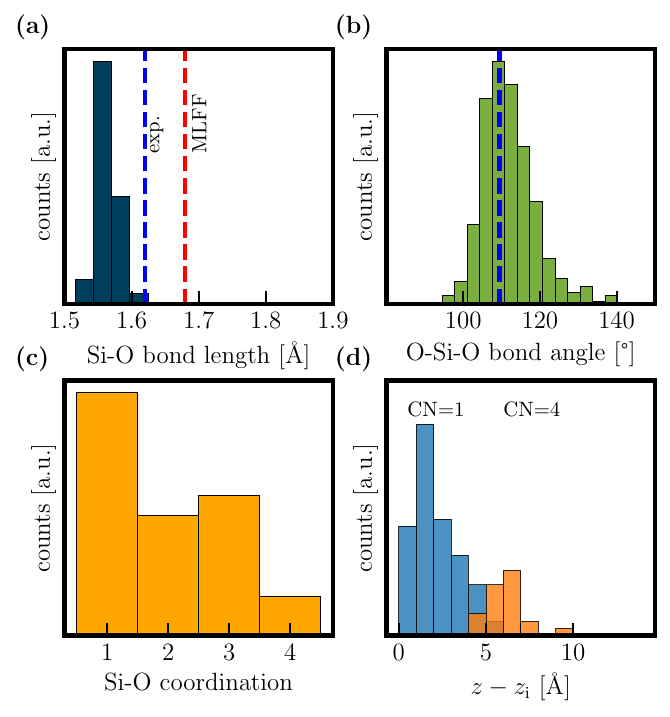}
\caption{Structural properties of a Si/SiO$_2$ interface generated by the reactive force field from~\cite{Nayir2019}. The plot is analogous to Fig.~\ref{fig:geometry}. For comparison, the experimental and averaged MLFF values for the bond lengths and the tetrahedral angles are shown as blue and red dashed lines, respectively. With respect to the MLFF and experiments~\cite{Diebold1999}, reaxFF slightly underestimates bond lengths while the tetrahedral angles are well captured. The properties of the interface, however, do not match experimental expectations, as O diffusion is overestimated by reaxFF. The number of four-fold O-coordinated Si is very low; on the other hand, there are many Si atoms with only one O neighbor. This results in a wide interface and a thin oxide layer.
}
\label{reaxFF_coordination}
\end{figure}

The growth kinetics, however, differ substantially between the two force fields. The number of oxygen molecules that dissolved at the surface is 15\,\% lower when using reaxFF. On the other hand, reaxxFF overestimates the diffusion of oxygen which leads to a low-density distribution of O atoms among the Si atoms in the crystal. The result is a very large interface (the transition region $d_\mathrm{IF}$ measures about 1\,nm and the oxide thickness $d_\mathrm{OX}$ is only 4\,\AA), with lower than expected Si--O coordination, see Fig.~\ref{reaxFF_coordination}. To summarize, the structures obtained by reaxFF are in strong contrast to the experimental findings of~\cite{Diebold1999, TEM2, Muller1999, Muller2001, Photo2}, while the MLFF -- similar to AIMD -- reproduces much more realistic interface structures.

\begin{table}[t]
\caption{Comparison between a classical reactive force field~\cite{Nayir2019}, AIMD simulations, the herein presented MLFF, and (if available) experimental values. 
Basic two- and three-body measures like Si--O bond lengths and O--Si--O angles are relatively well captured by all approaches, with slightly deviating values from reaxFF.
In terms of volumetric mass density, we compare the density in the interface region $\rho_\mathrm{IF}$ and in the oxide layer $\rho_\mathrm{OX}$.
Clear differences in the interface properties are indicated by the thickness of the interface $d_\mathrm{IF}$ and oxide thickness $d_\mathrm{OX}$ obtained after 1.4\,ns of dynamic oxidation at 1000\,K and 50\,bar.
}
\centering
\renewcommand{\arraystretch}{1.5}
\begin{tabular}{||l|c|c|c|c||} 
 \hline
 \multirow{2}{4em}{ } & 
 \multirow{2}{4em}{\centering reaxFF} &
 \multirow{2}{4em}{\centering AIMD\\\cite{Cvitko2023}} &
 \multirow{2}{4em}{\centering MLFF} &
 \multirow{2}{5em}{\centering experiment\\\cite{Diebold1999, TEM2, Muller1999}} \\
 & & & & \\ 
 \hline
 Si--O length~[\AA] & 1.57 & 1.66 & 1.68 & 1.62 \\ 
 O--Si--O angle [°] & 112.28 & 109 & 109.45 & 109.47 \\
 $\rho_\mathrm{IF}$ [g/cm$^3$] & 2.45 & 2.34 & 2.37 & 2.36--2.41 \\
 $\rho_\mathrm{OX}$ [g/cm$^3$] & 2.65 & 2.5 & 2.5 & 2.24--2.36 \\
 $d_\mathrm{IF}$ [nm] & 1 & 0.5 & 0.5 & 0.5 \\
 $d_\mathrm{OX}$ [nm] & 0.4 & - & 1 & - \\
 \hline
\end{tabular}
\label{comparison}
\end{table}

\subsection{Dangling bond density}
Finally, the last test we subject our model to is the determination of the dangling bond density resulting from the oxidation process.
In the most simple definition, a dangling bond corresponds to 
a missing neighbor, that is every Si atom with less than 4 and every O atom with less than two neighboring atoms is identified as a dangling bond.


By means of this simple analysis, we detect at least one dangling bond in 96\,\% of the obtained interface structures. Typically, these dangling bonds do not vanish if the structure is relaxed within a subsequent DFT optimization, as this requires the breaking of other bonds which is unlikely to happen.
On average, we find 2.9 dangling bonds in structures grown on a 1.5$\times1.5$nm$^2$ Si surface area, corresponding to a dangling bond density of 1.3/\,nm$^{2}$ ($130\times10^{12}$cm$^{-2}$).

Compared to the experimentally determined defect density of 0.05/\,nm$^{2}$ ($5\times10^{12}$cm$^{-2}$)~\cite{Stesmans_electrical_activity_1998}, our MLFF seems to overestimate the number of defects by almost two orders of magnitude. 
However, only ESR-active defects (depending on the applied bias and the location of the defect in the band gap) appear in the ESR measurement of ~\cite{Stesmans_electrical_activity_1998}.
Another important difference between simulation and experiment is the presence of hydrogen, which is unavoidable in reality, however, completely absent in our structures. Since atomic hydrogen passivates dangling bonds~\cite{Benton_H_pass_1980}, and therefore reduces the dangling bond density, a significant impact on the experimentally obtained value can not be ruled out. 
Furthermore, we suspect that the density of dangling bonds in the simulated structures could be reduced by equilibrating the structure at elevated temperatures for some \textmu{}s. 
Given these uncertainties, we conclude that our MLFF produces structures with significantly increased dangling bond density, although a direct comparison to values inferred from ESR measurements is not valid.

A detailed analysis of the dangling bonds in the simulated structures that goes beyond the simple coordination analysis above requires thorough investigations by means of DFT.
Within DFT, one can determine trap levels and relaxation energies~\cite{Wilhelmer_2022} and thereby investigate whether the defects are ESR-active.
We leave such extensions for the future.

\section{Conclusions}
In this study, we have introduced a Gaussian approximation potential (GAP)-based machine learning (ML) interatomic potential tailored for generating ultra-thin oxide layers on a Si substrate, including amorphous interfaces between silicon and its native oxide SiO$_2$.
The atomic structures are obtained by oxidizing an initially oxygen-free Si surface by means of molecular dynamics simulations.
Starting from entirely O-free Si surfaces, the (dry) thermal oxidation process is simulated by exposing the surface to an oxygen atmosphere at 50\,bar. 
The ML potential is capable of reproducing the intricacies of thermal oxidation in its initial stage (up to about 10\,nm oxide thickness), that is the transition from spontaneous O$_2$ surface reactions to molecular precursor mediated dissociations. Furthermore, the experimentally expected atomic configuration at the interface is reproduced accurately showing a 0.5\,nm thick transition layer with increasing O density between the Si crystal and the oxide layer.

The credibility of our simulation framework is validated by comparison of several geometric qualities of the interface structure with experimental values, which is remarkably good in terms of bond properties and atom coordination.
However, the obtained defect (or dangling bond) density exceeds the experimental values significantly, an effect that could be related to the relatively short simulation times (a few ns) compared to experimental conditions.

Furthermore, the applicability and usefulness of our ML model is highlighted by comparison to the highly successful reactive force field (reaxFF) in its most novel version~\cite{Nayir2019}. In this respect, the interfaces generated by the ML model are much closer to the experimental expectations. 

\section{Acknowledgments}
This project has received funding from the European Research Council (ERC) under grant agreement no. 101055379. The computational results presented have been achieved using the Vienna Scientific Cluster (VSC).

\appendix

\section{Descriptor parameters}
\label{app:descriptor}

The descriptors can be tuned by the user via a number of parameters as given in Tab.~\ref{SOAP_parameters}. The weight of each descriptor is controlled by $\delta$, $r_\mathrm{cut}$ is a cut-off radius which defines a sphere within which neighboring atoms are considered. $r_\mathrm{\Delta}$ is the cutoff transition
width, which defines the distance needed for the descriptor cut-
off to smoothly go to zero. $n_\mathrm{max}$ and $l_\mathrm{max}$ are the number of
angular and radial basis functions for the SOAP descriptor, respectively, and $\zeta$ is the power the kernel is raised to.

\begin{table}[t]
\caption{Parameters of the employed SOAP, two-body and three-body descriptors. The meaning of the parameters is given in the text (with detailed description for SOAP in~\cite{SOAP}).
}
\centering
\begin{tabular}{||l c c c||} 
 \hline
 Parameter & SOAP & Two-body & Three-body \\ [1ex] 
 \hline\hline
 $\delta$ & 0.4 & 4 & 1 \\ 
 $r_\mathrm{cut}$ & 5 & 4 & 3 \\
 $r_\mathrm{\Delta}$ & 1 & - & - \\
 $n_\mathrm{max}$ & 8 & - & - \\
 $l_\mathrm{max}$ & 4 & - & - \\
 $\zeta$ & 4 & - & - \\ [1ex] 
 \hline
\end{tabular}
\label{SOAP_parameters}
\end{table}

\section{Generation of the training dataset}
\label{app:trainingdata}
One of the main challenges when developing an MLFF is finding suitable training data. Among the problems that can be encountered are overfitting~\cite{Ying_2019}, data quality issues (incomplete or biased data), as well as imbalanced data (some classes of structures in the training dataset appear more frequently than others, resulting in a bias or poor performance).
A detailed overview of the data used for our MLFF is given in Tab.~\ref{training_structures}. The dataset contains single atoms, dimers, periodic bulk structures (crystalline and amorphous), surfaces and nanowires, as well as gaseous O$_2$. In a first step, we tried several data compositions and chose the MLFF that gave the best results (based on evaluation of the properties from Fig.~\ref{fig:geometry} and Tab.~\ref{comparison}). This initial ML model was then used to generate new structures that were recalculated by DFT. This additional data was then implemented into the training dataset from  Tab.~\ref{training_structures}. With this data, we obtain the final MLFF.

Our efforts to further improve the MLFF by including more data into the training dataset gave the opposite result: the performance decreased. From this, we conclude that the model is prone to overfitting. One of the ways to avoid overfitting is the so-called ``early-stopping strategy''~\cite{Ying_2019}. Since the MLFF already gave satisfying results in its second iteration, we decided to stop at this point.

\begin{table}[ht!]
\caption{Structures in the training dataset of the GAP force field. The type of structure is given along with the number of atoms in the structure as well as the number of individual structures. The associated parameters $\sigma_\mathrm{E}$ and $\sigma_\mathrm{F}$ represent the regularisation in the GAP corresponding to energies and forces, respectively~\cite{Klawohn_2023}.
}
\centering
\begin{tabular}{||l| c | c |c | c ||} 
 \hline
 Structure type & Number of & Number of & $\sigma_\mathrm{E}$ & $\sigma_\mathrm{F}$ \\  
                & atoms     & structures & & \\
 \hline\hline
 single atoms & 1 & 3 & 0.0001 & 0.001 \\
 Si dimers & 2 & 97 & 0.01 & 0.1 \\ 
 O dimers & 2 & 57 & 0.01 & 0.1 \\
 Si--O dimers & 2 & 23 & 0.01 & 0.1 \\
 Si--H dimers & 2 & 52 & 0.01 & 0.1 \\
 Si bulk & 192 & 90 & 0.002 & 0.02 \\
 SiO$_2$ bulk & 216 & 70 & 0.002 & 0.02 \\
 clean Si surface & 224 & 50 & 0.002 & 0.02 \\
 ox. Si surface & 232--5211 & 213 & 0.002 & 0.02 \\
 Si nanowire & 576--1680 & 100 & 0.002 & 0.02 \\
 ox. Si nanowire & 1682--2063 & 9 & 0.002 & 0.02 \\ 
 O$_2$ gas & 3--200 & 143 & 0.002 & 0.02 \\[1ex] 
 \hline
\end{tabular}
\label{training_structures}
\end{table}

\bibliographystyle{apsrev4-2}
\bibliography{my}

\end{document}